\documentclass{elsart}
\begin{document}
\runauthor{Shatner, Mushnik, Leshno, Solomon}
\begin{frontmatter}
\title{ A Continuous Time Asynchronous Model of 
the Stock Market;}
\subtitle{Beyond the LLS Model}
\author[neuron]{ M.~Shatner }
\author[phys]{L.~Mushnik}
\author[ba]{M.~Leshno}
\author[physall]{ S.~Solomon \thanksref{correspond}}
\address[neuron]{Center of Computational Neuroscience}
\address[phys]{ Racah Institute of Physics}
\address[ba]{ School of Business Administration, Mount Scopus }
\address[physall]{ Racah Institute of Physics \\ Givat Ram 91104 
\\ Hebrew University of Jerusalem}
\thanks[correspond]{Corresponding Author. E-mail: 
$sorin @ vms.huji.ac.il$}
\begin{abstract}
In order to simulate the complex phenomena manifested 
in stock markets, we introduce a continuous asynchronous
model in which millions of individual traders interact 
through a central orders matching
mechanism, just as it happens in real stock markets. 
Each trader has a unique
decision function, which allows him/ her to trade at any time, 
to react to external news, to respond to
price changes (or volume, volatility, etc.), 
and to consider the "fundamental price". 
As a simple example
we consider three "generic" decision functions, which correspond
to three trader profiles:  Noisy, Fundamentalist and Chartist. 
\end{abstract}
\end{frontmatter}
{\bf Introduction}

The microscopic simulation of the stock markets has been 
pursued extensively in the last decade by physicists and economists
in various collaborations
\cite{1,2,3,4,5,6,7}.

In particular Levy Levy and Solomon \cite{1,2,3,4,5} have introduced
a microscopic market model where the market price is established by the
condition of global market clearence.

The last market measurements show that much of the salient
information resides in the particular timing of the trades,
in their frequency and in the time correlations between the
trading frequency at various times \cite{0}.

To capture this crucial aspect of the market dynamics 
it is important that the new generation of market models
allow the traders to decide not only the price and volume
of their offers and bids but also their exact timing.

Moreover, all the events and time stages involving a transaction
(from the respective traders becoming interested in the market
until the posting of the transaction price
 by the market monitor) have to be
faithfully represented if the simulation is to reproduce the
trade-by-trade characteristics of the real market.

We introduce here a continous time
asynchronous model of the stock market, 
in which each trader's decision process takes place 
independently of the other traders'. 

Moreover, our model permits each trader to have a unique
decision process and to issue buy-sell orders at any time.

The current paper describes a stacks based algorithm
 which underlies the continuous asynchronous model (section
2.1), the market matching mechanism (sections 2.2-2.3) 
and three "generic" decision functions that
were used in current computer simulations of the model (section 3). 
A description of how we implement the model is given in section 4. 
Simulation results of the
model will be published in subsequent papers.
\section{The Market Mechanism}

A trader in our model is not continuously active, 
but rather "sleeps" most of the time.
Of course the time intervals when nothing happens are 
not actually spent by the computer waiting idly:
each operation/event in the virtual market world
is performed by the computer in the appropriate
time order and with the appropriate advancement
of the "market world clock" without actually waiting
for the a corresponding real time to pass. 

The trader
wakes up  at times which are the result of 
his own (previously taken) decisions:
e.g. the trader may decide to wake-up authomatically at a 
pre-defined time interval after the last order or 
in reaction to specific market events (news, price changes).
Once awake, the trader 
decides whether (s)he wishes to issue new Buy/Sell orders 
(see figure 1). 

When a trader first enters the market, (s)he 
is given an initial amount of cash and shares. In the
current version of the software implementing the model, 
a trader cannot have more than one order in the market 
at any given time (and cannot borrow or short-sale).

\subsection{The Stacks Algorithm}

In order to make the traders actions and decisions 
asynchronous and taking place at arbitrary continous
time, we implemented the market using a
set of stacks, that contain traders in different decision order stages (see figure 2):

\begin{itemize}
\item {\bf The "Dormant Stack"}  - 
 A trader's default state is dormant. 
Each trader defines the conditions on which 
(s)he should be waken up: 
(s)he can ask for instance to be waken up 
every time that a time K has passed since the last wake-up,
or (s)he can ask to be 
waken up when something happens in the market 
(news arrived to the market, price has changed by more than 
$5 \%$ etc.).  
External news events are given on an ordinal scale (e.g. -5 to 5), 
such that the trader can ask to be waken up only from a specific news
 level. 

\item {\bf The "Decision Stack"} -  Once the trader is awake, he is transferred to the "Decision
Stack". Each trader has an individual delay period, during which he stays in this stack (and
"examines" the market). Once the delay period is over, the trader can decide to issue a new order,
to change / delete an existing order (if it has not already been executed), or to go back to
sleep. He can also change his waking parameters. 
 An "order" in our model is actually a "LIMIT" order, i.e. a desired price and quantity of shares.
Future versions may also allow "MARKET" orders (i.e. buy/sell at the best market price). 

\item {\bf The "Transferred Orders Stack"} - When a trader decides to issue a new order, the order
is transferred to this stack before it enters the market (see 2.2 below). A transferred order is
delayed in this stack for a constant period of time (In future version each trader will have an
individual delay period, since some traders may be "closer" to the market). 
\end{itemize}

Note that when running a simulation of the model there is no relation between computer time intervals and simulated time interval, since computer operations are made as soon as possible without waiting.

\subsection{Matching Orders} 

Once an order leaves the "transferred orders" stack, 
it enters either the BUY stack or the SELL
stack of the market matching orders mechanism. The orders in these stacks are sorted according to
price. 

If the minimal price in the SELL stack is lower than the
maximal price in the BUY stack:

{\bf  MIN(SELL  Order  Price) 
$<$ MAX (BUY  Order   Price)  }

then a transaction of shares-for-cash 
will be executed between the issuers of these orders. Currently,
short selling of shares is not allowed in the model.

Of the two orders that were executed, the first that was issued to the market also determines the
new market price. If the two orders differ in size (number of shares), the number of shares in the
transaction equals the minimum size of the two orders. The order that was not fully executed is
returned to the "transferred orders" stack, with a decreased size.

In future versions of the program, we may limit the Market Orders Stacks in either size (worst
orders are cancelled) or time (orders with expiry times).

\subsection{The World Manager}

All this process of waking up the traders, moving their orders between 
the different stacks and
operating the market matching mechanism is performed by a central 
"World manager" (or Market Manager and a wider sense). 
Other than
being a technical aid to the process, 
the World manager has no influence on it whatsoever. In
future versions of the program, the World manager will also have the 
ability to introduce news to
the market.

\section{Traders Decision Functions}
\subsection{Traders Types}

When entering the "Decision Stack", a trader uses his individual decision function to process
inputs (current price, historical price quotes, news, fundamental price) and produce a decision
(Go back to sleep, Change waking parameters and issue an order). 

The decision function can be programmed to include any decision making algorithm. The current
version of the market simulation offers three types of pre-defined configurable decision
functions:
\begin{itemize}
\item {\bf "Random Trader"} - Issues buy/sell orders randomly in the proximity of the current
market price.
\item {\bf "Fundamentalist Trader"} - decides according to a "fundamental price", which follows a
random walk. 
\item {\bf "Chartist Trader"} - Uses simple technical analysis (In our case - curve fitting for
the last N price quotes).
\end{itemize}
Future versions of the model will include more complex "psychological" features of traders
\cite{9}, including the possibility to react to external news.

\subsection{Notation}

Following are descriptions of the three configurable decision functions. But first some notation:

When issuing an order, a trader has to decide on two parameters:

{\bf OrderP} = The price the trader sets for his order

{\bf OrderShares} = Shares to buy/sell (i.e. the size of the order)

The decision process also uses the following variables:

{\bf NewP} = This is the trader's best estimate for the market price in the next time tick.

{\bf MP(t), FP(t)} = 
Market price and fundamental price 
(respectively) at the current time.

{\bf CurrShares} = Current number of shares the trader owns.

{\bf CurrCash} = Current amount of cash the trader has.

{\bf PriceTick} = 
The market price is progressing in discrete ticks of size 
PriceTick (default: 1/32).

{\bf RAND[a,b]} = A random number in the interval [a,b].

{\bf RANDN[m,s]}  = A number drawn from a gaussian distribution with mean {\bf m} and standard
deviation {\bf s}.

\subsection{Random Trader}

In order to calculate the price for his order, a Random Trader chooses a random price near the
current market price: 

{\bf  OrderP = MP(t) + MP(t) * Sigma * RAND[-1,1] }

Where:

{\bf Sigma = 
Price variations around Market Price (configurable, default: 
$10\%$).}

Now the Random Trader chooses randomly the desired value of shares he wishes to hold after the
order is executed (in Currency value) :

{\bf InvestInShares = (OrderP * CurrShares + CurrCash) * RAND[0,1] }

He therefore needs to own this amount of shares:

{\bf InvestInShares / OrderP}

So he will buy /sell  (determined by the sign) the following number of shares:

{\bf  OrderShares = $|$ CurrShares - InvestInShares / OrderP $|$ }

3.4 Fundamentalist Trader

The basis for the fundamentalist trader's behavior is the fundamental price, FP, which is updated
every pre-defined update period:

 {\bf FP(t+1) = FP(t)  * ( RANDN[0,Sigma] +1 ) }

Where

{\bf Sigma }= Maximum price variations around FP (configurable, default: 0.4)

What the Fundamentalist Trader actually sees is a noisy Fundamental Price, {bf FP'(t) :
{\bf FP'(t) = FP(t) * ( 1+ RAND[-1,1]  * Sigma ) }

Note: If  {\bf FP'(t)}  $\leq$ 0, we set {\bf FP'(t) = FP(t)}.

The following ratio is used to determined how much the Fundamentalist is willing to trade (He is
more willing to trade as the difference between Market Price and his individual Fundamental Price
is higher):

{\bf RATIO = $|$ (MP(t) - FP'(t) ) / MP(t) $|$ }

Now the decision is easy: 

If {\bf FP'(t) $<$ MP(t) }, 

the trader wishes to sell, in a price as close as possible to the market price:

{\bf OrderP = MP(t) - PriceTick  }

If {\bf MP (t )= PriceTick} then {\bf OrderP=MP(t )}

{\bf OrderShares = RATIO * CurrShares. }

If {\bf FP'(t) $>$ MP(t), }  the trader wishes to buy shares, in the best price possible:

{\bf OrderP = MP(t) + PriceTick}

{\bf OrderShares = CurrCash * RATIO / OrderP }

\subsection{Chartist Trader}

A chartist trader has a memory span of N (default is 3). He extrapolates the next market price
({\bf NewP}) using a simple polynomial fit for the last N market price quotes he remembers (which
means that these price quotes need not be three consecutive market price quotes). 

Once {\bf NewP}  is determined:

If {\bf MP(t) $>$ NewP}   the trader wants to sell stocks:

{\bf OrderP = MP(t) - PriceTick}

{\bf OrderShares = CurrShares * (OrderP - NewP) / MP(t)}

If {\bf MP(t) $<$ NewP}   the trader wants to buy stocks:

{\bf OrderP = MP(t) + PriceTick}

{\bf OrderShares = CurrCash * (NewP - OrderP) / (MP(t )* OrderP ) }

\section{Implementing the Model}
Using the stacks system, our model can simulate a market with millions of traders acting continuously and asynchronously according to individual decision functions. There are no limitations on the number or types of decision functions that are used in the market.

We implemented the market model using a PC-based software. Here is a description of a typical market simulation: 
First, we need to introduce traders to the market. 
This is done by iteratively choosing a decision function and configuring it for a specific trader (see fig. 3). 
The decision functions to choose from are pre-programmed, and can vary from simple chartists to complicated "psychological" decision functions. 
For the sake of simplicity, we allow several traders to use the same configured decision function (forming a "group"), e.g. 10,000 traders that use a fundamentalist decision function with the same parameters. 
However, if the decision function contains random variables (as most do), each trader in a 
group will act independently in the market, with no correlation to the other members of the group (besides his decision function). 
The market in the current example (figure 3) consists of 6 different groups of fundamentalists (some are "richer", i.e. 
have more cash/shares when entering the market, some fluctuate more around the fundamental price), 3 groups of chartists ("richer", "longer memory", etc.) and two groups of random traders. 
Thus, a total of 161,014 traders were introduced to the market in 11 different groups.

After choosing the traders, the market simulation is initiated. The market form (see fig. 4) shows market parameters in real-time, including market price and data on the traders that are at the top of the BUY and SELL stacks. 
One can get simple graphs and statistics from the user interface, and a more thorough analysis using a direct link to matlab. 
Future versions of the software will also allow tracking a single trader's behavior in the market (current version allows it only manually). 
Figure 5 shows the price behavior of the simulation under the current settings.

\section{Concluding Remarks}
The continuous asynchronous model's main advantages are its flexibility, robustness and efficiency.  The model poses no limitation on the number of traders that enter the market, while allowing each trader to use a unique decision function. Moreover, the program implementing the model is fast enough to simulate a year of stock market activity with millions of traders in a matter of seconds.  
Results of the model will be published in subsequent papers.

\end{document}